\documentstyle[preprint,aps,epsf,floats]{revtex}

\begin{document}
\title{RNA Folding and Large N Matrix Theory}
\author{Henri Orland$^{1,2}$ and A.\ Zee$^{1}$}
\address{\vspace{.7cm}$^1$Institute for Theoretical Physics,\\
University of California, Santa Barbara, CA 93106, USA \\
$^{2}$Service de Physique Th\'eorique, CEA-Saclay,\\
91191 Gif-sur-Yvette Cedex, France}
\maketitle

\begin{abstract}
We formulate the RNA folding problem as an $N\times N$ matrix field theory.
This matrix formalism allows us to give a systematic classification of the
terms in the partition function according to their topological character.
The theory is set up in such a way that the limit $N\to \infty$ yields the
so-called secondary structure (Hartree theory). Tertiary
structure and pseudo-knots are obtained by calculating the $1/N^2$
corrections to the partition function. We propose a generalization of the
Hartree recursion relation to generate the tertiary structure.
\end{abstract}

\preprint{\tighten \vbox{\hbox{} }}

{\tighten
}



\newpage

\section{Introduction}

Over the last decade, RNA has transformed itself from being a relatively
minor player in the central dogma of Watson and Crick to being one of the
central players in molecular biology. Indeed, it has been recently
demonstrated that in addition to its ``information carrier'' role in
protein synthesis, some types of RNA's, known as ribozymes, have an
enzymatic activity which is crucial to the functioning of the cell \cite
{enzyme}. As a consequence of this new prominent role of RNA, the search for
the three dimensional structure of RNA has become an important problem
in biology. This view was expressed forcefully by Tinoco and Bustamante\cite
{old}.

As this paper is addressed to theoretical physicists, we begin with a
schematic review. A very thorough review on RNA folding can be found in ref. 
\cite{Higgs}.

RNA is a heteropolymer constructed out of a four-letter alphabet, $C,$ $G,$ $%
A,$ and $U$ (for the four bases or nucleotides cytosine, guanine, adenine,
and uracil). The length of an RNA chain ranges typically from 76 for tRNA to
a few thousand base pairs for mRNA. In solution, there is an attraction
between $C$ and $G$ and between $A$ and $U,$ with energies $\varepsilon
(C,G)\simeq -3$ kCal/mole and $\varepsilon (A,U)\simeq -2$ kCal/mole
respectively. There is also a weaker attraction between $G$ and $U$, with
energy $\varepsilon (G,U)\simeq -1$ kCal/mole. Note the correspondence 300 K 
$\simeq $ 0.6 kCal /mole $\simeq $ 1/40 eV.

Consider an RNA sequence $\{s\}=\{s_{1},s_{2},\cdots ,s_{L}\}$ (where $s_{i}$
takes on one of the four possible values $C,$ $G,$ $A,$ and $U$). For
example, we might be given the sequence $\{s\}=\{CCCGAAAUUCGUAG\cdots \}$.
The attraction between the nucleotides folds the RNA heteropolymer into a $%
3- $dimensional structure referred to as a shape. Biological functions
depend largely on the shape assumed by a particular RNA. Thus, the map from
sequence space to shape space is of great importance in molecular biology
and has been much discussed in the biophysical literature. As mentioned
above, this has been even more true since the discovery of the enzymatic activity
of some RNA.

In the molecular biology of biopolymers, it is conventional to define three
levels of structures. The primary structure is just the chemical sequence,
or sequence of nucleotides. The secondary structure is the local short-range
pairing of complementary bases, leading to segments of helices separated by
loops and bulges (``clover-leaf'' structure). Finally, the tertiary
structure is the spatial arrangement of these secondary motifs, in which the
loops and bulges themselves can partially pair, leading to the so-called
pseudo-knots (see fig.~1a).

\begin{figure}[tbh]
  \epsfxsize=0.5\linewidth
  \centerline{\hbox{ \epsffile{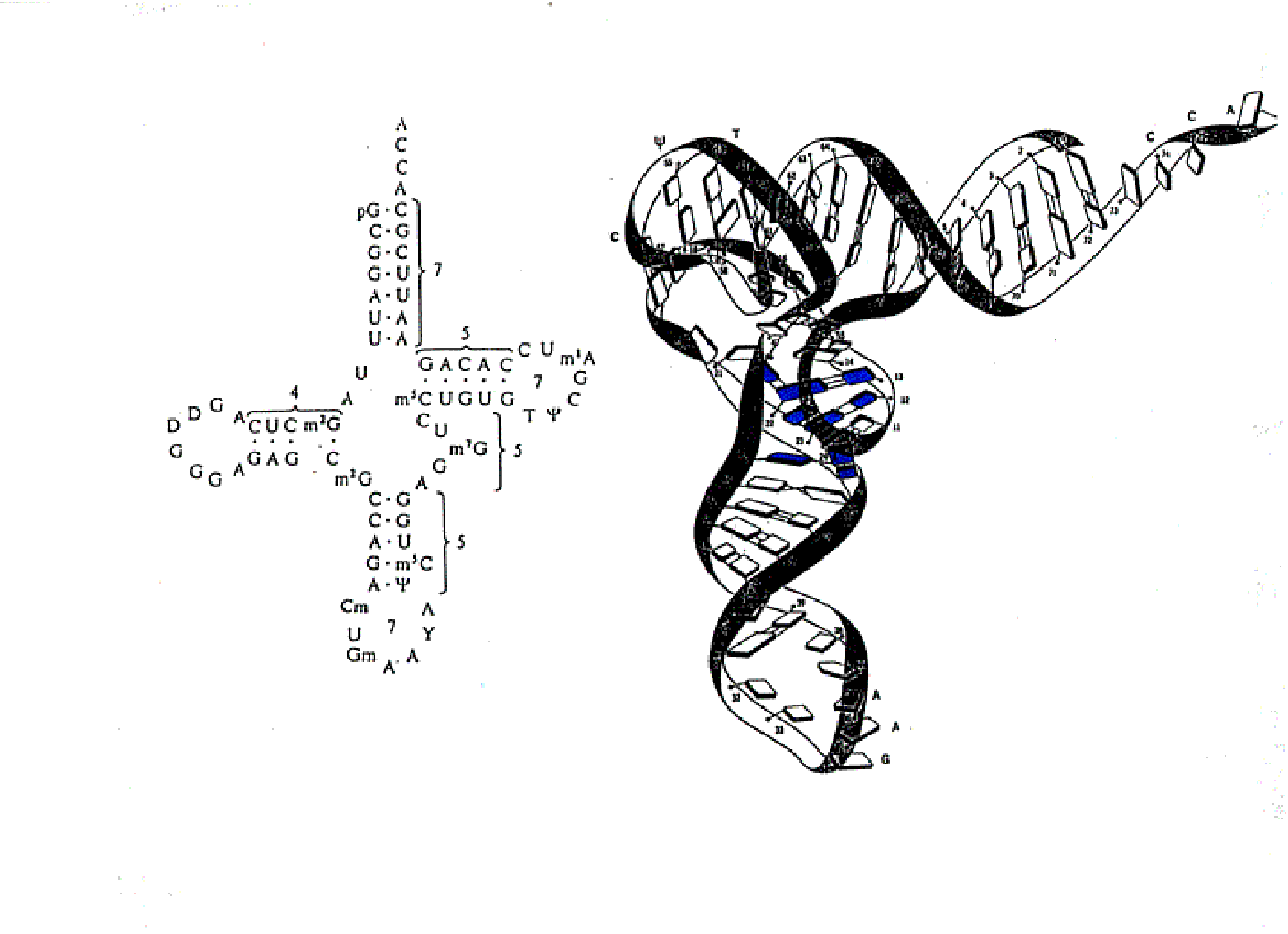} }}
Fig.~1a: Secondary (left) and tertiary (right) structure of a
tRNA.(From I. Tinoco, with permission.)
\end{figure}

An example of pseudo knot is the ``kissing hairpin'' Fig.~1b.

\begin{figure}[tbh]
  \epsfxsize=0.3\linewidth
  \centerline{\hbox{ \epsffile{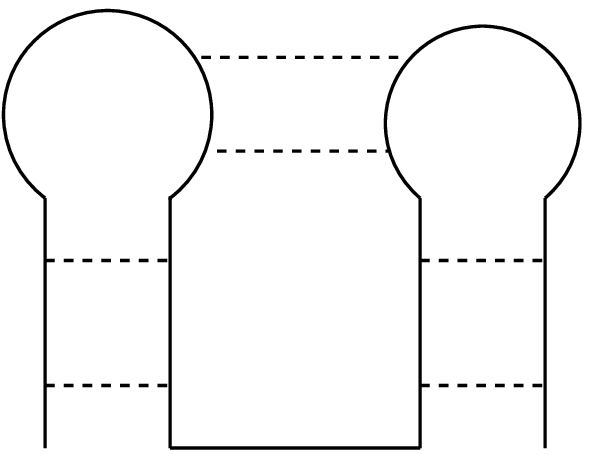} }}
  \centerline{ Fig.~1b: A ``kissing hairpin''.} 
\end{figure}

In contrast to the problem of protein folding \cite{creighton,garel}, RNA
folding is hierarchical in that its secondary structure is much more stable
than its tertiary structure, which can be treated as a perturbation\cite{old}%
. Experimentally, the two levels of folding (secondary and tertiary) can be
separated by varying the concentration of Mg$^{++}$ ions \cite{misra}. In
addition, the attractive force between nucleotides saturates. Once a given
nucleotide $C$ has paired with a nucleotide $G,$ it cannot be paired with
yet another $G.$ In contrast, the attraction between amino acids do not
saturate. Thus, the problem of RNA folding is considerably simpler than the
problem of protein folding.

The determination of secondary structure has reached a very high level of
sophistication based on dynamic programming algorithms \cite{Nussinov,Zuker,Vienna}.

The problem of RNA folding is clearly topological in flavor and is thus not
easily amenable to dynamic programming methods, although some 
algorithm has been proposed recently\cite{three}. On the other
hand, we know from the field theoretic literature that topological
considerations also play an important role in such subjects as matrix theory
or $M-$theory. In this paper, we propose that matrix theory may be useful to
the problem of RNA folding. We develop a matrix theoretic representation of
the topological aspect of RNA folding.

In section I, we formulate the RNA folding problem more precisely. In
section II, we show how it can be formulated as an $N\times N$ matrix field
theory. In section III, we show that the $N$ dependence of the field theory
can be made explicit in the functional integral formulation of the problem.
As a result, the natural way to compute the $1/N$ expansion is through a
steepest descent method which is described in section IV. As this expansion
is very complicated to perform at higher order, we resort in section V to
recursion relations which allow us to approximately incorporate the higher
order powers in $1/N$.

For a simple introduction to this work, one can go for instance to the website
http://online.itp.ucsb.edu/online/infobio01/zee/

\section{RNA Folding}

Given an RNA sequence $\{s\}=\{s_{1},s_{2},\cdots ,s_{L}\}$ of $L$ bases,
let us write down the partition function ${\cal Z}$ at temperature $1/\beta$%
. We will proceed in steps.

First, construct the matrix 
\begin{equation}
V_{ij}=e^{-\beta |\varepsilon
(s_{i},s_{j})| v(|\vec{r}_{i}-\vec{r}_{i}|)}\theta (|i-j|>4),i\neq j;V_{ii}=0.  \label{vij}
\end{equation}
where $\varepsilon (a,b)$ denotes the $4$ by $4$ real symmetric matrix
giving the attractive energy between nucleotides, $\varepsilon (A,U)$ etc.
We set the diagonal elements $V_{ii}$ to $0$ to indicate the fact that a
nucleotide does not attract itself. The Heaviside function $\theta (|i-j|>4)$
incorporates the fact that the RNA molecule is not infinitely flexible and
we cannot pair nucleotides separated by less than 4 sites. 
The attractive potential can be taken to be $v(r)=-w\theta (R-r)$ with $w$
and $R$ the strength and range of the attraction respectively.

Now construct

\begin{equation}
Z=1+{\sum_{<ij>}}V_{ij}+{\sum_{<ijkl>}}V_{ij}V_{kl}+\cdots +{\sum_{<ijkl>}}
V_{ik}V_{jl}+\cdots  \label{Z}
\end{equation}
where $<ij>$ denotes all pairs with $j>i$, $<ijkl>$ all quadruplets with $%
l>k>j>i,$ and so on. Then the partition function is given by 
\begin{equation}
{\cal Z=}\int \prod_{k=1}^{L}d^{3}\vec{r}_{k}\prod_{i=1}^{L-1}f(|\vec{r}%
_{i+1}-\vec{r}_{i}|)\,\,Z  \label{Z1}
\end{equation}
The function $f(r)$ can be taken to be, for example, $\delta (r-l)$ for a
model in which the nucleotides are connected along the RNA heteropolymer by
rigid rods of length $l$, or $e^{-(r-l)^{2}/6\sigma ^{2}}$for a model with
elastic rods. Note that the saturation of the hydrogen bond has been
incorporated by the requirement $l>k>j>i,$ and so on. Once the nucleotide at 
$i$ has interacted with the nucleotide at $j$ it cannot interact with the
nucleotide at $k$ . Note that in (\ref{Z}), only the enthalpy and
combinatorics of pairings are included. The integration over the atomic
coordinates in (\ref{Z1}) accounts for the actual topological feasibility of
a given pairing and also for the entropic factor associated with loop
formation.

Biologists are interested in the folded configuration essentially at room
temperature. Since room temperature is substantially less than the melting
temperature (of order $80^{0}$C, in other words, the characteristic energy
scale of the problem), we want to determine the ground state configuration
of the RNA heteropolymer. In other words, once we have obtained $Z$ we would
like to extract the term in $Z$ that dominates as $\beta \varepsilon $ tends
to infinity in (\ref{vij}).

We have given a simplified quantitative framework for the RNA folding
problem. From a chemical point of view, it would be appropriate to include
also the stacking energies of couples of complementary base pairs, instead
of energies of single pairs of bases. However, in the following, we will
stick with the latter. We will also concentrate on the
evaluation of the ``pairing'' partition function (\ref{Z}). We expect that
the various effects we have ignored, such as stacking 
, etc..., can be
added later as ``bells and whistles'' to our approach.
The stacking energies for instance can be taken into account by utilizing a $%
16\times16$ interaction matrix between pairs of bases instead of the $4\times4$
matrix $\varepsilon(s_i,s_j)$ we use here.

\section{Matrix Theory}

What is the connection with matrix theory?

Consider pulling apart the folded RNA structure given in fig.~2a. 

\bigskip 
\begin{figure}[tbh]
  \epsfxsize=0.3\linewidth
  \centerline{\hbox{ \epsffile{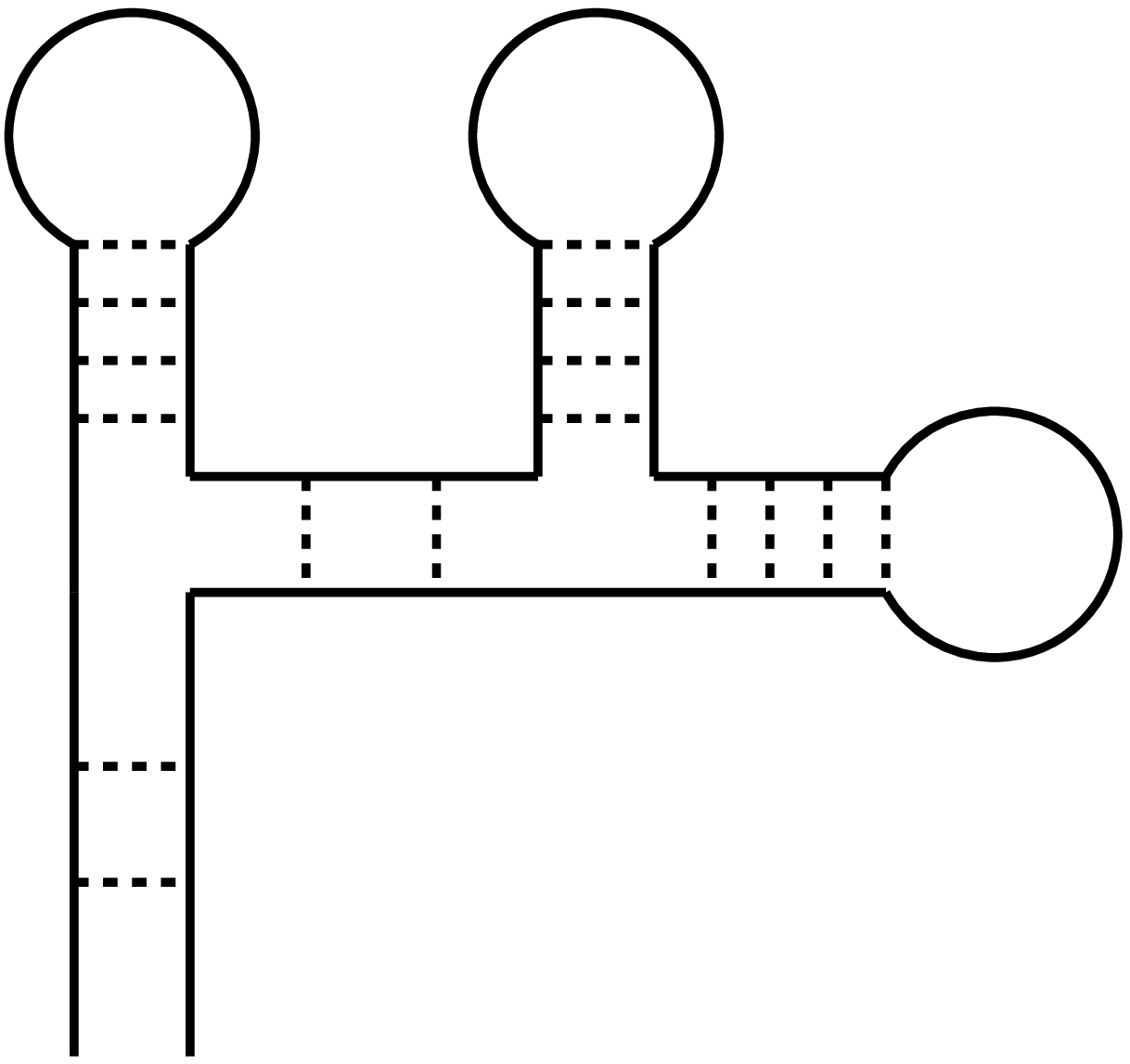} }}
  \centerline  { Fig.~2a: Representation of the secondary structure of an RNA.}
\end{figure}

We obtain the structure of fig.~2b which to physicists are reminiscent of
Feynman diagrams in a variety of subjects: matrix theory, quantum
chromodynamics, and so on.

\begin{figure}[tbh]
  \epsfxsize=0.3\linewidth
  \centerline{\hbox{ \epsffile{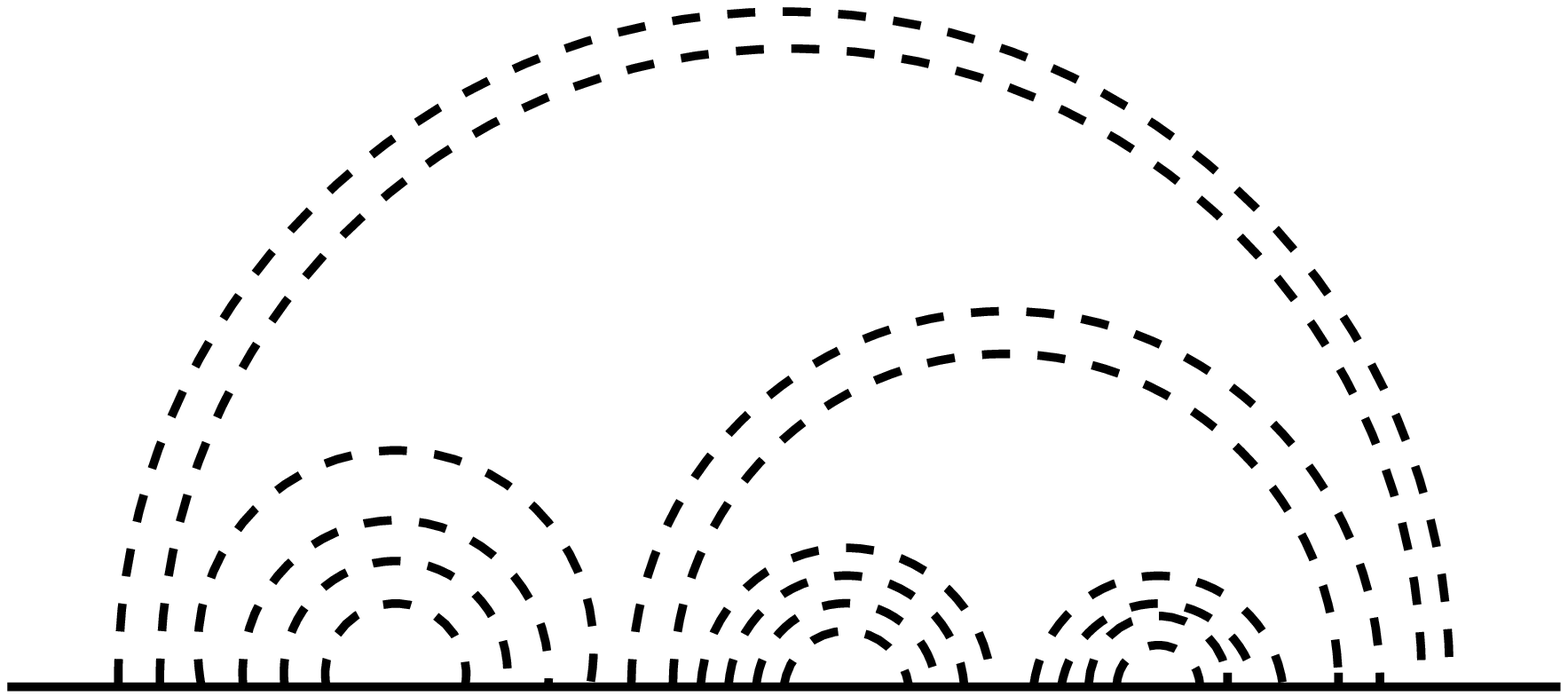} }}
  \centerline  { Fig.~2b: Representation of the same RNA stretched.}
\end{figure}

\bigskip For the sake of definiteness, let us borrow the terminology of
quantum chromodynamics. The dotted lines are known as gluon propagators, and
the solid line as a quark propagator. The secondary structure corresponds to
diagrams in which the gluon lines do not cross over each other, while the
tertiary structure corresponds to diagrams in which the gluon lines do cross.

The crucial observation, originally made by 't Hooft \cite{witten}, is that 
there is a systematic relation between the topology of a graph
and its corresponding power of $1/N^2$.
For instance, planar diagrams are of order $1/N^0$, and diagrams in
which gluon 
lines cross are of higher order.
We merely
have to go to the large $N$ expansion, and the diagrams are classified by
powers of $1/N^2$. Note that a somewhat similar formulation in terms of
matrix theory has been used for the meander problem \cite{Dif_Gui}.

Consider the quantity 
\begin{equation}
Z(1,L)=\frac{1}{A(L)}\int \prod_{k=1}^{L}d\varphi _{k}e^{-\frac{N}{2}{%
\sum_{ij}}(V^{-1})_{ij}\,{\rm tr}(\varphi _{i}\varphi _{j})}\frac{1}{N} %
\mathop{\rm tr}\prod_{l=1}^{L}(1+\varphi _{l})  \label{matrep}
\end{equation}
Here $\varphi _{i}$ $(i=1,$ $\cdots ,$ $L)$ denote $L$ independent $N$ by $N$
Hermitian matrices and ${\Pi _{l}}(1+\varphi _{l})$ represents the ordered
matrix product $(1+\varphi _{1})(1+\varphi _{2})\cdots (1+\varphi _{L}).$
All matrix products will be understood as ordered in this paper. The
normalization factor $A(L)$ is defined by 
\begin{equation}
A(L)=\int \prod_{k=1}^{L}d\varphi _{k}e^{-\frac{N}{2}{\sum_{ij}}(V^{-1})_{ij}%
{\rm tr}(\varphi _{i}\varphi _{j})}
\end{equation}

Let us refer to the row and column indices $a$ and $b$ of the matrices $%
(\varphi _{i})_{a}^{b}$ as color indices, with $a,b=1,2,\cdots ,N$. The
matrix integral (\ref{matrep}) defines a matrix theory with $L$ matrices. We
can either think of it as a Gaussian theory with a complicated observable $%
\frac{1}{N}{\rm tr}{\Pi _{l}}(1+\varphi _{l})$, or alternatively, by raising 
$\frac{1}{N}{\rm tr}{\Pi _{l}}(1+\varphi _{l})=e^{\log [\frac{1}{N}{\rm tr}{%
\Pi _{l}}(1+\varphi _{l})]}$ into the exponent, as a complicated matrix
theory with the action $(\frac{N}{2}{\sum_{ij}}(V^{-1})_{ij}{\rm tr}(\varphi
_{i}\varphi _{j})-\log [\frac{1}{N}{\rm tr}{\Pi _{l}}(1+\varphi _{l})])$.
Another trivial remark is that we can effectively remove $\frac{1}{N}{\rm tr}
$ from (\ref{matrep}).

The important remark is that the matrix theory \cite{Bre_Zee} defined by (%
\ref{matrep}) has the same topological structure as 't Hooft's large $N$
quantum chromodynamics. There are $L$ types of gluons, and the gluon
propagators are given by $\frac{1}{N}V_{ij}.$ As in large $N$ quantum
chromodynamics, each gluon propagator is associated with a factor of $\frac{1%
}{N}$ and each color loop is associated with a factor of $N.$ The reader
familiar with matrix theory or large $N$ quantum chromodynamics can see
immediately that the Gaussian matrix integral (\ref{matrep}) evaluates
precisely to the infinite series 
\begin{equation}
Z(1,L)=1+{\sum_{<ij>}}V_{ij}+{\sum_{<ijkl>}}V_{ij}V_{kl}+\cdots +\frac{1}{%
N^{2}}{\sum_{<ijkl>}}V_{ik}V_{jl}+\cdots   \label{z1L}
\end{equation}
Some ``typical'' terms in this series correspond to the diagrams in fig.~3.

\begin{figure}[tbh]
  \epsfxsize=0.5\linewidth
  \centerline{\hbox{ \epsffile{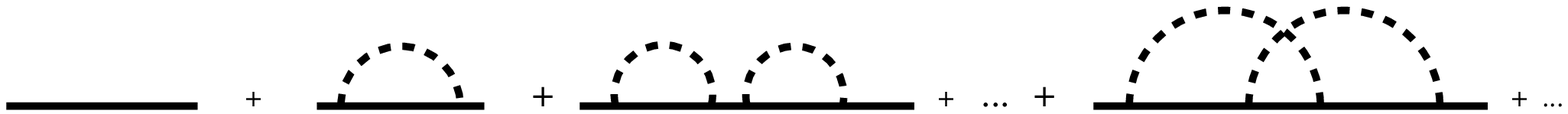} }}
  \centerline  { Fig.~3: Graphical representation of a few terms of the partition
function.}
\end{figure}

This differs from (\ref{Z}) only in that the terms with different 
topological character
are now classified by inverse powers of $\frac{1}{N^{2}}$. Thus,
the use of the large $N$ expansion allows us to separate out the tertiary
structure, represented in (\ref{z1L}) for example by the term $\frac{1}{N^{2}%
}{\sum_{<ijkl>}}V_{ik}V_{jl},$ from the secondary structure.

Note that the ordered product ${\Pi _{l}}(1+\varphi _{l})$ ensures that the
diagonal elements $V_{ii}$ of the matrix $V$ do not appear in $Z(1,L)$. We
have nevertheless already set $V_{ii}$ to 0.

The program proposed in this paper is thus to evaluate $Z(1,L)$ with $V$ an
arbitrary matrix. Once $Z(1,L)$ is known we can then insert it into (\ref{Z1})
to evaluate ${\cal Z}$. The parameter $\frac{1}{N}$ serves as a convenient
marker to distinguish the tertiary structure from the secondary structure.
What we offer here is a systematic way of generating refinements to the
calculation of $Z,$ and hence the free energy $F,$ to any desired accuracy
in a well controlled approximation.

Since in $Z(1,L)$ the quantities $1$ and $L$ represent arbitrary labels we
can just as well define 
\begin{equation}
Z(m,n)=\frac{1}{A(m,n)}\int \prod_{k=1}^{L}d\varphi _{k}e^{-\frac{N}{2}%
\Sigma _{i,j=m}^{n}(V^{-1})_{ij}{\rm tr}(\varphi _{i}\varphi _{j})}\frac{1}{N%
}{\rm tr}\prod_{l=m}^{n}(1+\varphi _{l})  \label{Zmn}
\end{equation}
where again the normalization is given by

\[
A(m,n)=\int \prod_{k=m}^{n}d\varphi _{k}e^{-\frac{N}{2}{\sum_{ij=m}^{n}}
(V^{-1})_{i,j}{\rm tr}(\varphi _{i}\varphi _{j})} 
\]

As we shall see in the following, we will construct recursion relations to
evaluate (\ref{Zmn}) approximately. These recursion relations can be easily
programmed to calculate the free energy of the RNA chain.

\section{Large N}

\bigskip In the matrix representation (\ref{matrep}) $N$ appears implicitly
in the size of the matrices $\varphi _{i}.$ In order to study the large $N$
limit, we need to extract the $N$ dependence explicitly, for which we have
developed the following method. Define $G_{l^{\prime }l}$ by $G_{l^{\prime
}l}=\prod_{i=l}^{l^{\prime }-1}(1+\varphi _{i})$ for $l^{\prime }-1\geq l,$ $%
G_{l-k,l}=0$ for all $k>0,$ and $G_{ll}=1.$ Then $G_{l^{\prime }l}$
satisfies the equation 
\begin{equation}
G_{l^{\prime }l}-(1+\varphi _{l^{\prime }-1})G_{l^{\prime }-1,l}=\delta
_{l^{\prime }l}
\end{equation}
Thus, if we define $M_{l^{\prime }l}=\delta _{l^{\prime }l}-(1+\varphi
_{l^{\prime }-1})\delta _{l^{\prime }-1,l}$ \ then we see that $%
G_{ll^{\prime }}$ is the inverse of the matrix $M_{ll^{\prime }}$ and thus 
\begin{eqnarray}
Z(1,L) &=&\frac{1}{A(L)}\int {\Pi _{k}}d\varphi _{k}e^{-N\frac{1}{2}{%
\sum_{ij}}(V^{-1})_{ij}{\rm tr}(\varphi _{i}\varphi _{j})}M_{L+1,1}^{-1}
\label{zrep} \\
&=&\frac{1}{A(L)}\int {\Pi _{k}}d\varphi _{k}e^{-N\frac{1}{2}{\sum_{ij}}%
(V^{-1})_{ij}{\rm tr}(\varphi _{i}\varphi _{j})}\int {\Pi }_{l}\,d\psi
_{l}^{\ast }d\psi _{l}e^{-\psi _{l}^{\ast }M_{ll^{\prime }}\psi _{l^{\prime
}}}\psi _{L+1}\psi _{1}^{\ast }
\end{eqnarray}
We have used the standard representation of the inverse of a matrix by an
integral over Grassmanian fermionic variables $\psi _{l}$ and $\psi
_{l}^{\ast }.$ Note the felicitous fact that $\det M=\int d\psi ^{\ast
}d\psi e^{-\psi ^{\ast }M\psi }=1$ which allows us to write (\ref{zrep})
without a denominator.

To compactify this representation of $Z$ further we introduce $%
M(h)_{ij}=M_{ij}+h\delta _{i,1}\delta _{j,L+1}$ and write 
\begin{equation}
Z(1,L)=\frac{1}{N}\frac{\partial }{\partial h}\frac{1}{A(L)}\int {\Pi _{k}}%
d\varphi _{k}e^{-N\frac{1}{2}{\sum_{ij}}(V^{-1})_{ij}{\rm tr}(\varphi
_{i}\varphi _{j})}\int d\psi ^{\ast }d\psi e^{-\psi ^{\ast }M(h)\psi }
\end{equation}
Henceforth, it is understood that after differentiation with respect to $h$
we set $h$ to 0.

We can now perform the Gaussian integration over $\varphi _{k}$, thus obtaining 
\begin{equation}
Z(1,L)=\frac{1}{N}\frac{\partial }{\partial h}\int d\psi ^{*}d\psi
e^{-S_{0}(\psi ^{*},\psi )-S_{1}(\psi ^{*},\psi )}
\end{equation}
with the free fermion action 
\begin{equation}
S_{0}(\psi ^{*},\psi )=\sum_{j}(\psi _{j}^{*}-\psi _{j+1}^{*})\psi
_{j}+h\psi _{1}^{*}\psi _{L+1}
\end{equation}
and the interacting fermion action 
\begin{equation}
S_{1}(\psi ^{*},\psi )=-\frac{1}{2N}\sum_{j,j^{\prime }}\sum_{a,b}\psi
_{a,j+1}^{*}\psi _{j}^{b}V_{jj^{\prime }}\psi _{b,j^{\prime }+1}^{*}\psi
_{j^{\prime }}^{a}  \label{intS}
\end{equation}
Note that in (\ref{intS}) we have displayed the color indices $a$ and $b$
explicitly.

We next rewrite 
\begin{equation}
S_{1}(\psi ^{\ast },\psi )=+\frac{1}{2N}\sum_{j,j^{\prime }}K_{jj^{\prime
}}K_{j^{\prime }j}=\frac{1}{2N}\mathop{\rm tr}K^{2}
\end{equation}
in terms of the color singlet variable 
\begin{equation}
K_{jj^{\prime }}=\sum_{a}(V_{jj^{\prime }})^{\frac{1}{2}}\psi _{a,j+1}^{\ast
}\psi _{j^{\prime }}^{a}
\end{equation}
Now use the Gaussian representation 
\begin{equation}
e^{-\frac{1}{2N}{\rm tr}K^{2}}=\frac{1}{C}\int dA e^{-\frac{N}{2}{\rm tr}%
A^{2}+i{\rm tr}AK}
\end{equation}
with the normalization factor $C=\int dA e^{-\frac{N}{2}{\rm tr}A^{2}}.$
Note that even though $K$ is complex we can take $A$ to be hermitean.
(Equivalently, the anti-hermitean part of $A$ drops out.) Putting it
together we obtain 
\begin{equation}
Z(1,L)=\frac{1}{N}\frac{\partial }{\partial h}\frac{1}{C}\int dA e^{-\frac{N%
}{2}{\rm tr}A^{2}}\int d\psi ^{\ast }d\psi e^{-\sum_{ij}\sum_{a}\psi
_{a,i}^{\ast }M_{ij}\psi _{j}^{a}}  \label{repp}
\end{equation}
where 
\begin{equation}
M_{ij}=\delta _{ij}-\delta _{i,j+1}+h\delta _{i,1}\delta
_{j,L+1}+i(V_{i-1,j})^{\frac{1}{2}}A_{i-1,j}
\end{equation}
or in matrix form

\begin{equation}
M_{L}=\left( 
\begin{array}{lllllll}
1 & 0 & 0 & \cdot & \cdot & 0 & h \\ 
-1 & 1+a_{12} & a_{13} & \cdot & \cdot & a_{1L} & 0 \\ 
a_{12}^{\ast } & -1 & 1+a_{23} & \cdot & \cdot & a_{2L} & 0 \\ 
\cdot & \cdot & \cdot & \cdot & \cdot & \cdot & \cdot \\ 
\cdot & \cdot & \cdot & \cdot & \cdot & \cdot & \cdot \\ 
\cdot & \cdot & \cdot & \cdot & -1 & 1+a_{L-1L} & 0 \\ 
a_{1L}^{\ast } & a_{2L}^{\ast } & \cdot & a_{L-2L}^{\ast } & a_{L-1L}^{\ast }
& -1 & 1
\end{array}
\right)  \label{matrix}
\end{equation}
where we have used the convenient notation 
\begin{eqnarray}
i \sqrt{V_{ij}} \ A_{ij}&=& a_{ij} \ \ \ {\rm for}\ \ i < j  \nonumber \\
i \sqrt{V_{ij}} \ A_{ij}&=& a^{*}_{ji} \ \ \ {\rm for}\ \ j < i
\end{eqnarray}

The point of these manipulations is that in (\ref{repp}) we have now
isolated the color index $a$ so that the integral over $\psi ^{\ast }$ and $%
\psi $ factors into $N$ copies of the same integral, thus giving 
\begin{equation}
Z(1,L)=\frac{1}{N}\frac{\partial }{\partial h}\frac{1}{C}\int dA e^{-\frac{N%
}{2}{\rm tr}A^{2}}(\det M(A))^{N}=\frac{1}{N}\frac{\partial }{\partial h}%
\frac{1}{C}\int dA e^{-\frac{N}{2}{\rm tr}A^{2}+N{\rm tr}\log M(A)}
\label{res1}
\end{equation}
At this point, we can differentiate with respect to $h$ and set $h$ to 0,
obtaining the alternative form 
\begin{equation}
Z(1,L)=\frac{1}{C}\int dA e^{-\frac{N}{2}{\rm tr}A^{2}+N{\rm tr}\log
M(A)}M^{-1}(A)_{L+1,1}  \label{res2}
\end{equation}
In this expression, 
\begin{equation}
M_{ij}=\delta _{ij}-\delta _{i,j+1}+i(V_{i-1,j})^{\frac{1}{2}}A_{i-1,j}
\end{equation}
Let us introduce the action 
\begin{equation}
S(A)=\frac{1}{2}{\rm tr}A^{2}-{\rm tr}\log M(A)
\end{equation}
and define the average of an ``observable'' $O$ by 
\begin{equation}
<O>=\frac{1}{C}\int dAe^{-NS(A)}O
\end{equation}
(Note the non-standard normalization used here.) Then, our result can be
summarized elegantly as 
\begin{equation}
Z(1,L)=<M^{-1}(A)_{L+1,1}>  \label{bbbbb}
\end{equation}

At this point, as remarked earlier, we note that the quantity $Z(1,L)$ can
obviously be generalized to $Z(i,j)$: after all, the site labels $1$ and $L$
are arbitrary. Then we have the appealing result that 
\begin{equation}
Z(i,j)=<M^{-1}(A)_{j+1,i}>\text{ for }j>i  \label{zij}
\end{equation}

It is also useful to introduce the free action 
\begin{equation}
S_{0}(A)=\frac{1}{2}{\rm tr}A^{2}
\end{equation}
and to define 
\begin{equation}
<O>_{0}=\frac{1}{C}\int dA e^{-NS_{0}(A)}O.
\end{equation}
Then we can also write our result as 
\begin{equation}
Z(1,L)=\frac{1}{N}\frac{\partial }{\partial h}<(\det M(A))^{N}>_{0}
\label{b0}
\end{equation}

Remarkably, it turns out that we will need both the representations (\ref
{bbbbb}) and (\ref{b0}) later in a single calculation.

Incidentally, our formulation of the RNA folding problem can be immediately
adapted to the marriage problem (or bipartite matching problem) \cite{orland}
\cite{mezard}\cite{nieuwenhuizen}\cite{zhang}, one of the classic problems
in combinatorial optimization. We will mention only the simplest version
here. Label $L$ (with $L$ even) individuals by the index $i=1,$ $\cdots ,L$
where the individual is male for $i$ odd and female for $i$ even. Define a
matrix $V_{ij}=\frac{1}{2}(1-(-1)^{i+j})e^{-\beta E_{ij}}$ where $E_{ij}$
represents the energy cost of a marriage between $i$ and $j$ and its
negative provides a measure of happiness. Referring back to (\ref{Z}) we see
that we want to extract in $Z(1,L)$ all the terms with $L/2$ powers of $V,$
for example the term $V_{14}V_{38}\cdots V_{L-1,2}=e^{-\beta E_{Total}}$
with $E_{Total}=E_{14}+E_{38}\cdots E_{L-1,2}.$ Since we now want to include
possible crossings in the Feynman diagram language we can set the number of
colors $N$ to 1. Thus, from (\ref{res1}), we have immediately $Z(1,L)=\frac{%
\partial }{\partial h}\frac{1}{C}\int dAe^{-\frac{1}{2}{\rm tr}A^{2}}\det
M(A).$ Referring to (\ref{matrix}) we see that the differentiation with
respect to $h$ and setting $h$ to 0 amounts to replacing the $L+1$ by $L+1$
matrix $M(A)$ by the $L$ by $L$ matrix obtained by deleting the first row
and last column. Furthermore, since we want the terms with $L/2$ powers of $%
V,$ that is, with $L$ powers of $V^{\frac{1}{2}},$ we can set the $1$'s and $%
-1$'s in this matrix to 0. Denoting the resulting matrix by ${\cal M} (A)$,
we obtain the following representation for the marriage problem 
\begin{equation}
Z_{m}(L)=\frac{1}{C}\int dAe^{-\frac{1}{2}{\rm tr}A^{2}}\det {\cal M} (A)
\end{equation}

Clearly, the representation given here can be generalized in a number of
directions, for example, by including individuals who remain single.

It is easy to see how this representation works: the Gaussian integration
insures that in $\det {\cal M} (A)$ only the appropriate terms are picked
out.

\section{Steepest Descent\label{steep}}

The fact that we have been able to display explicitly the $N$ dependence is
crucial and allows us in principle to carry the $1/N$ expansion to any
order. The standard strategy to evaluate integrals such as (\ref{res2}) is
of course to use the method of steepest descent (\cite{BIPZ},\cite{bbb}).

To leading order the steepest descent approximation is easy enough to carry
out. The stationary point is determined by $\frac{\delta S(A)}{\delta A}=0,$
that is 
\begin{equation}
A_{lk}=i(V_{lk})^{\frac{1}{2}}G_{l-1,k+1}  \label{aij}
\end{equation}
where we find it useful to define 
\begin{equation}
G_{ij}=(M^{-1})_{i+1,j}  \label{G}
\end{equation}
Notice that with this definition $G_{ij}$ is defined for $i$ from $0$ to $%
L-1 $ and for $j$ from $2$ to $L+1.$ The identity $%
\sum_{j}M_{ij}(M^{-1})_{jk}=\delta _{ik}$ can now be written as 
\begin{equation}
G_{i+1,k}-G_{ik}-\sum_{j}V_{i+1,j}G_{i,j+1}G_{j-1,k}=\delta _{i+2,k}
\label{grec}
\end{equation}
Referring to (\ref{res2}) and (\ref{zij}) we see that to leading order in
steepest descent, $Z(i,j)$ is just $M^{-1}(A)_{j+1,i}=G_{ji}$ \ evaluated at
the stationary point.

This equation (\ref{grec}) has already been written down in the literature 
\cite{deGennes,Nussinov,Zuker,Vienna,Wat,McC,Bun_Hwa,Lub_Nel,Mon_Mez} and is known as the ``Hartree
approximation''. It has the obvious interpretation (see fig.~4) that to
lowest order the additive effect of including one extra nucleotide labelled
by $L+1$ to the RNA heteropolymer can be described by pairing that
nucleotide to the nucleotide labeled by $j,$ which separates the
heteropolymer into two segments, one from $1$ to $j$ and the other from $j+1$
to $L+1$. We then sum over all possible $j$ of course.

\begin{figure}[tbh]
  \epsfxsize=0.5\linewidth
  \centerline{\hbox{ \epsffile{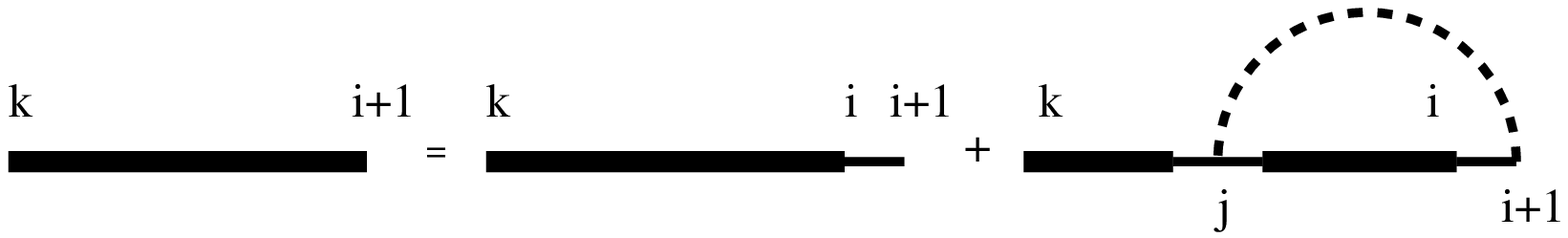} }}
Fig.~4: Graphical representation of the Hartree recursion
relation. The thick line represents the propagator $G$.
\end{figure}

In principle, steepest descent gives a systematic expansion of $Z(1,L)$ to
any desired power of $\frac{1}{N}$ by expanding the exponent and the
observable around the saddle-point. In the present context, this implies
that the full three dimensional structure of the RNA can be obtained by
expanding around the secondary structure. In particular, the higher order
terms do not disrupt the secondary structure, but merely add new
interactions, in addition to the existing secondary pairing. This is in
marked contrast with protein folding, where it is known that there is a
strong correlation between secondary and tertiary structure.

In practice, however, it proves to be quite tedious to calculate the $\frac{1%
}{N^{2}}$ terms explicitly. In the integral in (\ref{res2}) we are now to
replace $A_{ij}\ $ by $A_{ij}+x_{ij}/\sqrt{N}$ where $A_{ij}$ is determined
by (\ref{aij}) and (\ref{grec}). A straightforward calculation shows that

\begin{eqnarray}
Z(1,L) &=&\int dx\,\exp \left( -\frac{1}{2}{\rm tr\,}x^{2}-\frac{1}{2}{\rm tr%
}\left( M^{-1}c\right) ^{2}-\sum_{p=3}^{\infty }\frac{(-1)^{p}}{pN^{p/2-1}}%
{\rm tr}\left( M\,^{-1}c\right) ^{p}\right)  \nonumber \\
&\times &\left\{ \left( 1+\sum_{p=1}^{\infty }\frac{\left( -1\right) ^{p}}{%
N^{p/2}}\left( M^{-1}c\right) ^{p}\right) M^{-1}\right\} _{L+1,1}
\label{zex}
\end{eqnarray}
where $M^{-1}$ is related to $G$ through equation (\ref{G}), and $%
c_{ll^{\prime }}=\sqrt{V_{l-1,l^{\prime }}}\,x_{l-1,l^{\prime }}$ . The
systematic corrections to $Z$ are obtained by expanding (\ref{zex}) in
powers of $1/N^{1/2}$. By symmetry, no half-integer powers of $N$ remain in
the expansion of $Z$.

The first thing to evaluate is the propagator of the fluctuation fields $%
x_{ij}$. This is just the inverse of the kernel of the quadratic form
appearing in the exponent of (\ref{zex}). This propagator $\Delta _{ij,kl}$ is in fact
a scattering amplitude and
satisfies a form of the Bethe-Salpeter equation\cite{bethe}

\begin{equation}
\Delta _{kl,mn}=\delta _{km}\delta
_{nl}\,+\sum_{ij}V_{kl}^{1/2}V_{ij}^{1/2}G_{k-1,i+1}G_{j-1,l+1}\Delta
_{ij,mn}  \label{BS}
\end{equation}
where $G$ are the Hartree propagators (\ref{grec}). In fig.~5, we show
a graphical representation of this recursion equation, as well as the
series of graphs it resums. It is clear that this equation resums all
the possible ladder (or rainbow) diagrams to this order.

\begin{figure}[tbh]
  \epsfxsize=0.7\linewidth
  \centerline{\hbox{ \epsffile{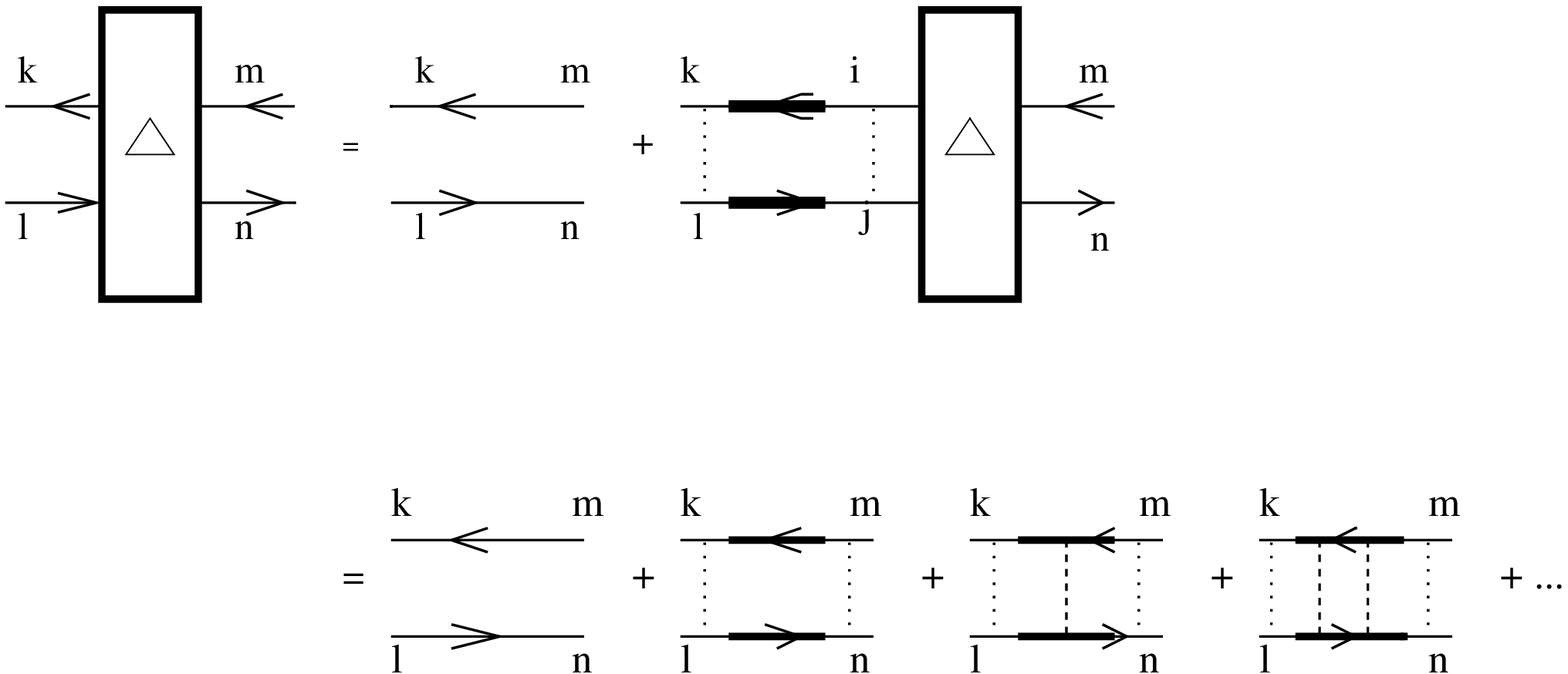} }}
Fig.~5: Graphical representation of the Bethe-Salpeter recursion
relation. The dotted lines represent factors $\sqrt{V_{ij}}$
while the dashed lines represent factors  $V_{ij}$.
The solid thick lines represent Hartree propagators $G$.
The Hartree propagators being directed, the arrows denote the
direction of increasing spatial index.
\end{figure}

This equation is to be solved for the particular sequence studied.
The scattering amplitude $\Delta$ defines the contractions of the $x$ fields, and
thus its knowledge allows us in principle to calculate (\ref{zex}) to any
order. Note that as usual in field theory, only contractions which are
linked to the operator that we calculate are to be included. (This reduces
considerably the number of contraction).

A fairly simple calculation allows us to show that the $1/N$ correction
vanishes identically (see appendix A). This result appears true by drawing a
few graphs, but this gives an algebraic proof.

It is easy to see that we have to expand (\ref{zex}) to $O(x^{6})$ in order
to calculate the free energy to order $\frac{1}{N^{2}}$. The
calculation, although cumbersome, is straightforward. The free energy
reads

\begin{eqnarray}
\label{correction}
Z(1,L) &=& G_{1L} \nonumber \\
&+&{1 \over N^2}< \biggl\{ \biggl( -{1\over 5} B_1 T_5 + {1 \over 12} 
B_1 T_3 T_4 - {1 \over 162} B_1 T_3^3  \nonumber \\
&-&{1\over 4} B_2 T_4 + {1 \over 18} B_2 T_3^2 -{1 \over 3} 
B_3 T_3 +B_4 \biggr) M^{-1} \biggr \}_{L+1,1} >
\end{eqnarray}
where we have used the notation

\begin{eqnarray}
D_{mn}&=&\sum_{m'} M^{-1}_{mm'} \sqrt{V_{m'-1,n}}\ x_{m'-1,n} \nonumber \\
\left(B_p \right)_{kl} &=& \left( D^p \right) _{kl} \nonumber \\
T_p &=& {\rm Tr} B_p
\end{eqnarray}

In (\ref{correction}), the bracket means that the Wick theorem should
be applied to contract the fields $x_{ll'}$ which appear in this expression, their
contraction being given by the kernel $\Delta$.

The calculation of the correction to the free energy is possible
numerically for not too long RNA sequences. Work in this direction is
in progress.

Because of the complexity of the (exact) order $1/N^2$
obtained in this approach, we found it
simpler to generalize the Hartree recursion equation to incorporate some
residual interactions between the loops and bulges. 

\bigskip

\section{Recursion Approach}

Two approaches can be used to derive recursion relations for the partition
functions. One is detailed in the following, whereas the other one is
described in appendix B.

A possible approach is to take the expression in (\ref{b0}) 
\begin{equation}
Z(1,L)=\frac{1}{N}\frac{\partial }{\partial h}<(\det M(A))^{N}>_{0}
\end{equation}
and try to relate $Z(1,L+1)$ to $Z(1,L)$. In other words, we would like to
relate $<(\det M_{L+1}(A))^{N}>$ to $<(\det M_{L}(A))^{N}>$ where the
subscript on $M$ keeps track of the different matrices in the discussion.
Note that $M_{L}$ is an $L+1$ by $L+1$ matrix. Explicitly, as noted before,
the $L+2$ by $L+2$ matrix $M_{L+1}$ has the form 
\begin{equation}
M_{L+1}=\left( 
\begin{array}{lllllll}
1 & 0 & 0 & \cdot & \cdot & 0 & h \\ 
-1 & 1+a_{12} & a_{23} & \cdot & \cdot & b_{1} & 0 \\ 
a_{12}^{\ast } & -1 & \cdot & \cdot & \cdot & b_{2} & 0 \\ 
\cdot & \cdot & \cdot & \cdot & \cdot & \cdot & \cdot \\ 
\cdot & \cdot & \cdot & \cdot & \cdot & \cdot & \cdot \\ 
\cdot & \cdot & \cdot & \cdot & -1 & 1+b_{L} & 0 \\ 
b_{1}^{\ast } & b_{2}^{\ast } & \cdot & b_{L-1}^{\ast } & b_{L}^{\ast } & -1
& 1
\end{array}
\right)  \label{m}
\end{equation}
where for convenience we have denoted

\begin{eqnarray}
i \sqrt{V_{ij}}\ A_{ij}&=& a_{ij} \ \ \ {\text for}\ \ i < j \le L  \nonumber
\\
i \sqrt{V_{i,L+1}}\ A_{i,L+1}&=& b_{i} \ \ \ {\text for}\ \ i \le L 
\nonumber \\
i \sqrt{V_{ij}}\ A_{ij}&=& a^{*}_{ji} \ \ \ {\text for}\ \ j < i \le L 
\nonumber \\
i \sqrt{V_{L+1,j}}\ A_{L+1,j}&=& b^*_j \ \ \ {\text for}\ \ j \le L
\end{eqnarray}

Our strategy is to first perform the Gaussian integration over the $b_{j}$'s
in evaluating $<(\det M_{L+1}(A))^{N}>$, keeping in mind that we need the
terms of order $h$. This method of integrating out a row and a column has
also been used in random matrix theory\cite{bz}.

We briefly outline the procedure. Write $M_{L+1}=M_{L+1}(b=0)+B$ where $B$
is the matrix extracted from (\ref{m}) upon keeping only the entries which
depend on the $b$'s and $b^{*}$'s. Expand $(\det M_{L+1}(A))^{N}$ in powers
of $B$ and then perform the Gaussian average over the $b$'s and $b^{*}$'s,
using $<b_{i}b_{j}^{*}>=\frac{1}{N}\delta _{ij}V_{j,L+1}$. After some
arithmetic, we obtain 
\begin{eqnarray}
Z(1,L+1)&=&Z(1,L)  \nonumber \\
&+& \sum_{j=1}^{L}V_{j,L+1}<(\det M)^{N}[(\frac{\partial }{\partial h}%
M_{j,L+2}^{-1})M_{L+1,j+1}^{-1}-\frac{1}{N}(\frac{\partial }{\partial h}%
M_{L+1,L+2}^{-1})M_{j,j+1}^{-1}]>_{0}  \label{blll}
\end{eqnarray}
We have suppressed the subscript $L+1$ on the matrix $M$ on the right hand
side. It is understood that this expression is to be evaluated at $h=0.$
Noting that the matrix $\frac{\partial M}{\partial h}$ is particularly
simple and that ($M^{-1})_{L+2,L+2}=1$, we find that 
\begin{equation}
Z(1,L+1)=Z(1,L)+\sum_{j=1}^{L}V_{j,L+1}<M_{L+1,j+1}^{-1}M_{j,1}^{-1}-\frac{1%
}{N}M_{L+1,1}^{-1}M_{j,j+1}^{-1}>
\end{equation}

Using the definition of the connected expectation value 
\[
<AB> = <A><B>+<AB>_C 
\]
we note, as is well-known, that the connected part is of order $1/N^2$ (\cite
{witten}) and we can thus write 
\begin{eqnarray}
Z(1,L+1) &=&Z(1,L)+\sum_{j=1}^{L}V_{j,L+1}<M_{L+1,j+1}^{-1}><M_{j,1}^{-1}>
\label{recur} \\
+\sum_{j=1}^{L}V_{j,L+1} &<&M_{L+1,j+1}^{-1}M_{j,1}^{-1}>_{C} \\
-\frac{1}{N}\sum_{j=1}^{L}V_{j,L+1} &<&M_{L+1,1}^{-1}M_{j,j+1}^{-1}>_{C}
\end{eqnarray}

Recalling (\ref{zij}) we recognize that the quantities $<M_{L+1,j+1}^{-1}>$
and $<M_{j,1}^{-1}>$ appearing in the second term on the right hand side of (%
\ref{recur}) are nothing but $Z(j+1,L+1)$ and $Z(1,j)$ respectively. Thus,
if we keep only the first two terms on the right hand side of (\ref{recur})
we obtain the closed recursion relation 
\begin{equation}
Z(1,L+1)\simeq Z(1,L)+\sum_{j=1}^{L}V_{j,L+1}Z(1,j)Z(j+1,L+1)
\label{hartree}
\end{equation}

This is precisely the recursion relation in the Hartree approximation (\ref
{hartree}) mentioned earlier.

As announced in the introduction, the formulation given here offers a
systematic way to go beyond the Hartree approximation. We expect the third
and fourth term on the right hand side of (\ref{recur}), when evaluated to
leading order in $\frac{1}{N}$ to give the corrections of order $\frac{1}{%
N^{2}}.$ It is intriguing then that the superficially similar objects $%
<M_{L+1,j+1}^{-1}M_{j,1}^{-1}>_{C}$ and $<M_{L+1,1}^{-1}M_{j,j+1}^{-1}>_{C}$
must be of order $\frac{1}{N^{2}}$ and order $\frac{1}{N}$ respectively. We
note however that a ``backward-propagating object'' which we define as $%
M_{jk}^{-1}$ with $k>j$ makes its first appearance in $%
<M_{L+1,1}^{-1}M_{j,j+1}^{-1}>_{C}$. All other terms in (\ref{recur})
involve only forward-propagating objects.

We can of course calculate (\ref{recur}) explicitly for small $L$ in order
to check our formulation and the order of the various terms in $\frac{1}{N}$
. The result for $L=5$ is shown graphically in fig.~6.

\begin{figure}[tbh]
  \epsfxsize=0.3\linewidth
  \centerline{\hbox{ \epsffile{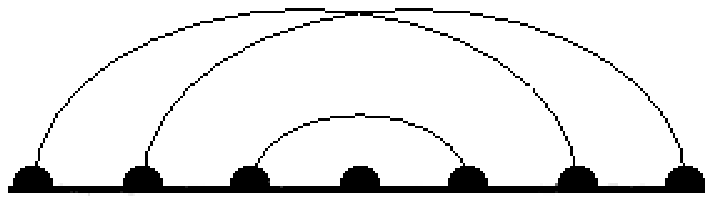} 
  \epsfxsize=0.3\linewidth  
  \epsffile{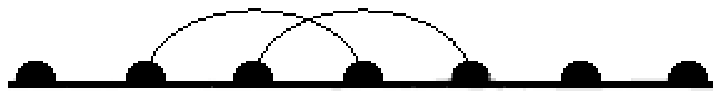} 
  \epsfxsize=0.3\linewidth  
  \epsffile{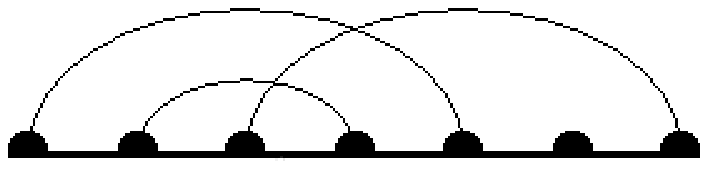}}}
  \centerline  { Fig.~6: A few graphs corresponding to the $1/N^2$ term. }
\end{figure}

\subsection{Recursion Relation}

While the recursion relation (\ref{recur}) has an appealing structure, we
are not able to evaluate the two objects $<M_{L+1,j+1}^{-1}M_{j,1}^{-1}>_{C}$
and $<M_{L+1,1}^{-1}M_{j,j+1}^{-1}>_{C}$ and express them in a simple form.
Neither should we be able to do that. Our experience in field theory, for
example the Dyson-Schwinger equation in quantum electrodynamics, indicates
that recursion relations generically do not close: new objects appear in the
right hand side. There is no reason why $<M_{L+1,j+1}^{-1}M_{j,1}^{-1}>_{C}$
should be expressible in terms of $<M_{ik}^{-1}>$. New objects,
corresponding to vertex functions in field theory, must appear.

Fortunately, we can inspect the set of Feynman diagrams to obtain a
recursion relation for $Z(i,j)$. We propose the following recursion
relation. Given $Z(i,j)$ for all $i$ and $j$ satisfying $j-i\leq L-1,$ we
obtain $Z(i,j)$ for all $i$ and $j$ satisfying $j-i\leq L$ as follows.

First, define $Z^{1PI}(i,j)$ as the one-particle irreducible (1PI)
part of $Z(i,j)$, that is
the sum of all those diagrams in $Z(i,j)$ that do not fall apart into two
separate pieces when a quark propagator is cut. Some examples are shown in
fig.~7a. 

\begin{figure}[tbh]
  \epsfxsize=0.5\linewidth
  \centerline{\hbox{ \epsffile{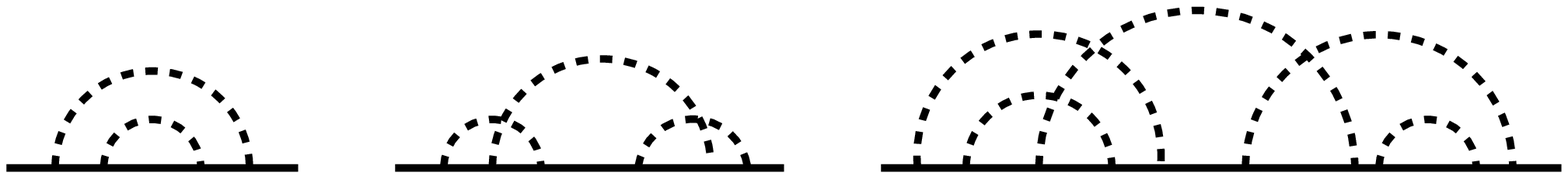} }}
  \centerline  { Fig.~7a: A few one particle irreducible graphs. }
\end{figure}

In fig.~7b, we show a different representation of the third graph of fig.~7a

\begin{figure}[tbh]
  \epsfxsize=0.2\linewidth
  \centerline{\hbox{ \epsffile{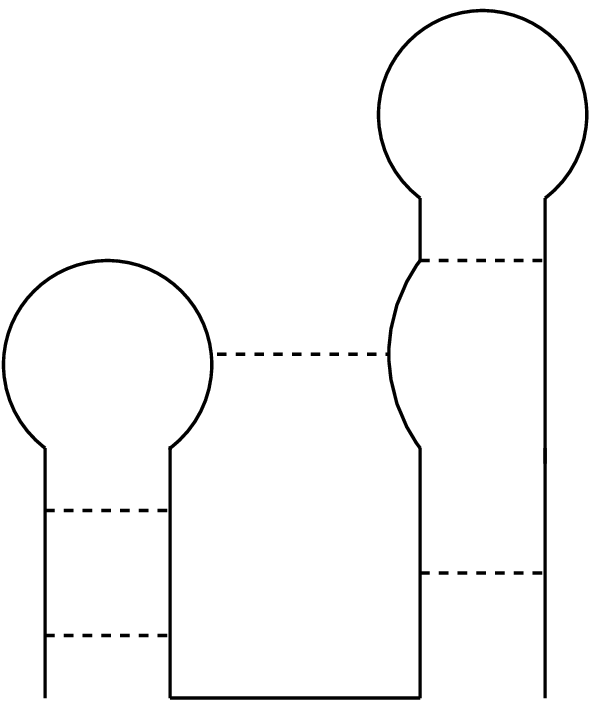} }}
  \centerline  { Fig.~7b: The third graph of fig.~7a}
\end{figure}

The concept of, and the necessity of introducing, one-particle
irreducibility is of course the same here as in field theory such as
quantum electrodynamics.

Second, define the vertex function $\Gamma _{mn}^{j}$ for $n>j>m$ by 
\begin{equation}
\label{gamma}
\Gamma _{mn}^{j}=\frac{1}{N^{2}}[1-\sum_{k\ne j}V_{jk}\frac{\partial }{
\partial V_{jk}}](Z^{1PI}(m,n)-1)
\end{equation}
Using the language of quantum chromodynamics, this equation is actually easy
to describe in words. The vertex function $\Gamma _{mn}^{j}$ describes a
quark propagating from $m$ to $n$ and interacting with a gluon at site $j$.
The operator $[1-\sum_{k \ne j}V_{jk}\frac{\partial }{\partial V_{jk}}]$
simply insures that there is not already a gluon attached to the site $j.$
See fig.~7a. The relation between $\Gamma _{mn}^{j}$ and $Z^{1PI}(m,n)$ has
the same form as the Ward identity in quantum electrodynamics.

Since we want to calculate $Z$ to order $1/N^2$, according to
eq.(\ref{gamma}) we need to keep only the 1PI diagrams of order $1$ 
in $Z^{1PI}(m,n)$ and in $\Gamma_{mn}^{j}$. 
These are just the 1PI Hartree diagrams, i.e. the
sum of all rainbow diagrams.
Note that $\Gamma_{mn}^{j}$ is simply related to the Bethe-Salpeter
scattering amplitude $\Delta _{kl,mn}$ (\ref{BS}) by
\begin{equation}
\Gamma_{mn}^{j} = \sqrt{V_{mn}} \sum_{k,l} \sqrt{V_{kl}} \Delta_{kl,mn}
\end{equation}
where the summation over $k,l$ is restricted to $m<k<j<l<n$. This
relation is represented graphically in fig.~8a

\begin{figure}[tbh]
  \epsfxsize=0.3\linewidth
  \centerline{\hbox{ \epsffile{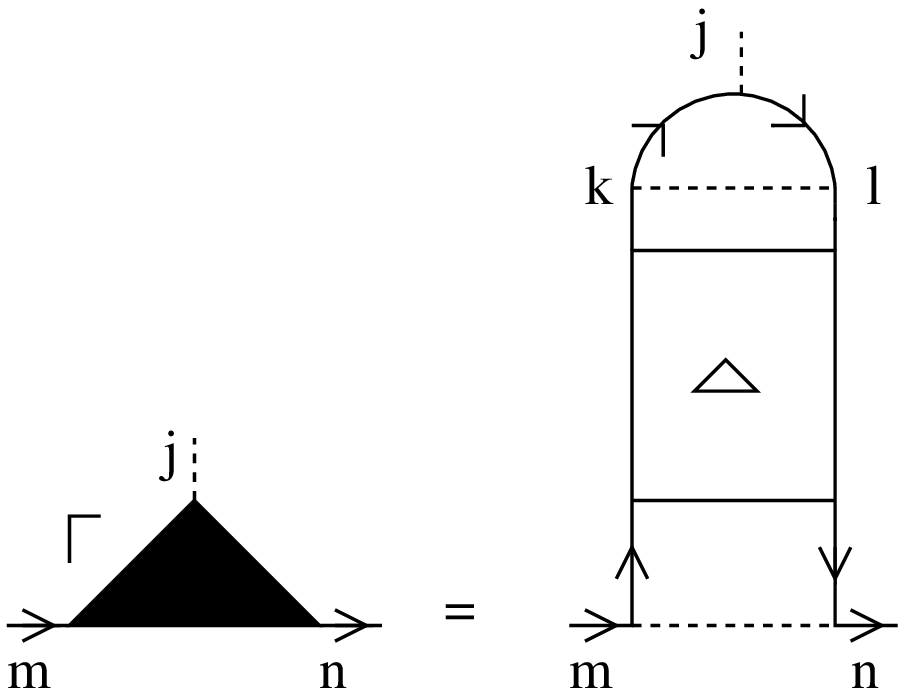} }} 
Fig.~8a: Graphical representation of the relation between
the vertex function $\Gamma_{mn}^{j}$ and the scattering amplitude
$\Delta _{kl,mn}$.
\end{figure}
 
Third, we calculate for $k+1>i$%
\begin{eqnarray}
\label{recursion2}
Z(i,k+1) &=&Z(i,k)+\sum_{j=1}^{k}V_{j,k+1}Z(i,j-1)Z(j+1,k)  \nonumber
\\
&&+\sum_{j=1}^{k}V_{j,k+1}\sum_{m,n}Z(i,m-1)\Gamma _{mn}^{j}Z(n+1,k)
\end{eqnarray}
with the boundary condition $Z(i,i)=1$, $Z(i,i-1)=1,$ and $Z(1,0)=1$. The
meaning of this equation is expressed graphically in fig.~8b.

\begin{figure}[tbh]
  \epsfxsize=0.5\linewidth
  \centerline{\hbox{ \epsffile{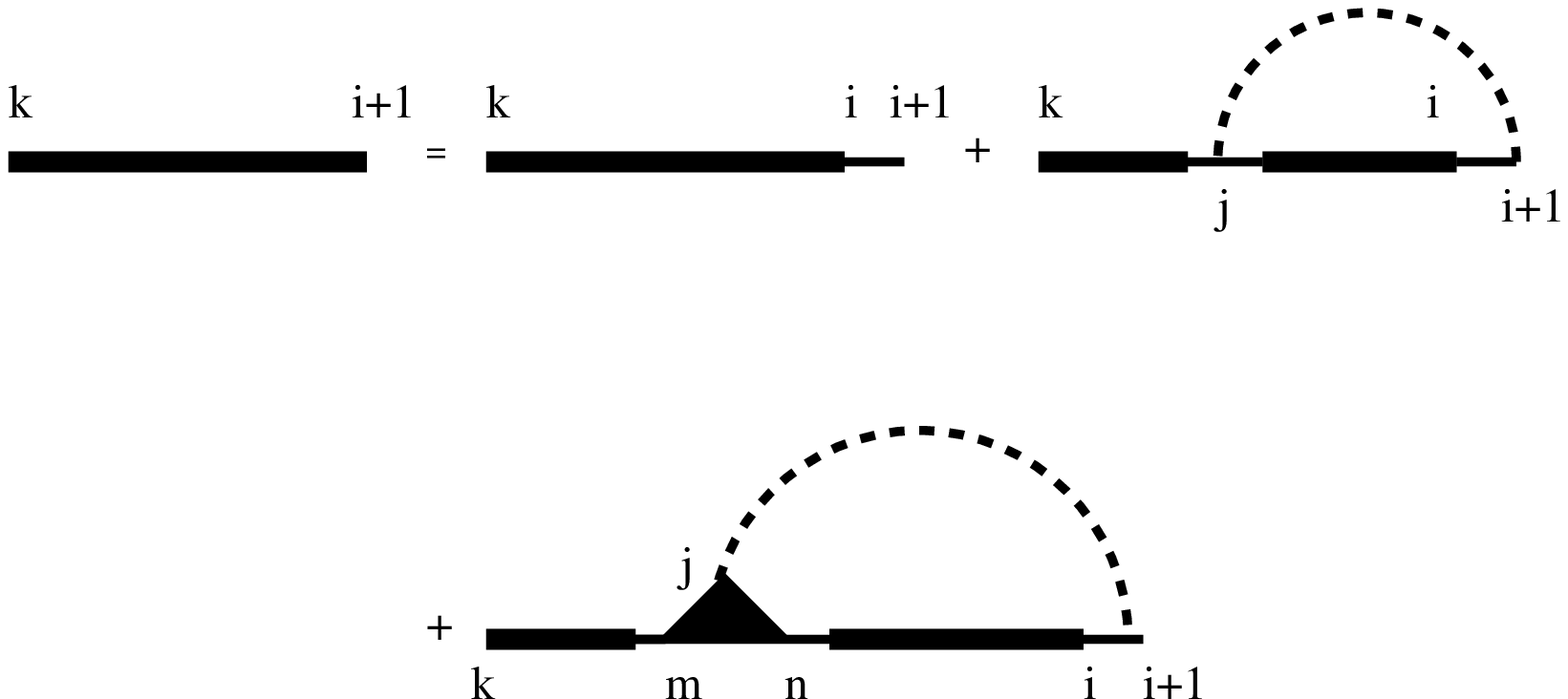} }} 
Fig.~8b: Graphical representation of the recursion equation to
order $1/N^2$. The black triangle represents the vertex function $\Gamma
_{mn}^{j}$.
\end{figure}

We have checked this equation explicitly for $L$ up to 6. A graph
generated to order $1/N^2$ is displayed in fig.~9.

\begin{figure}[tbh]
  \epsfxsize=0.3\linewidth
  \centerline{\hbox{ \epsffile{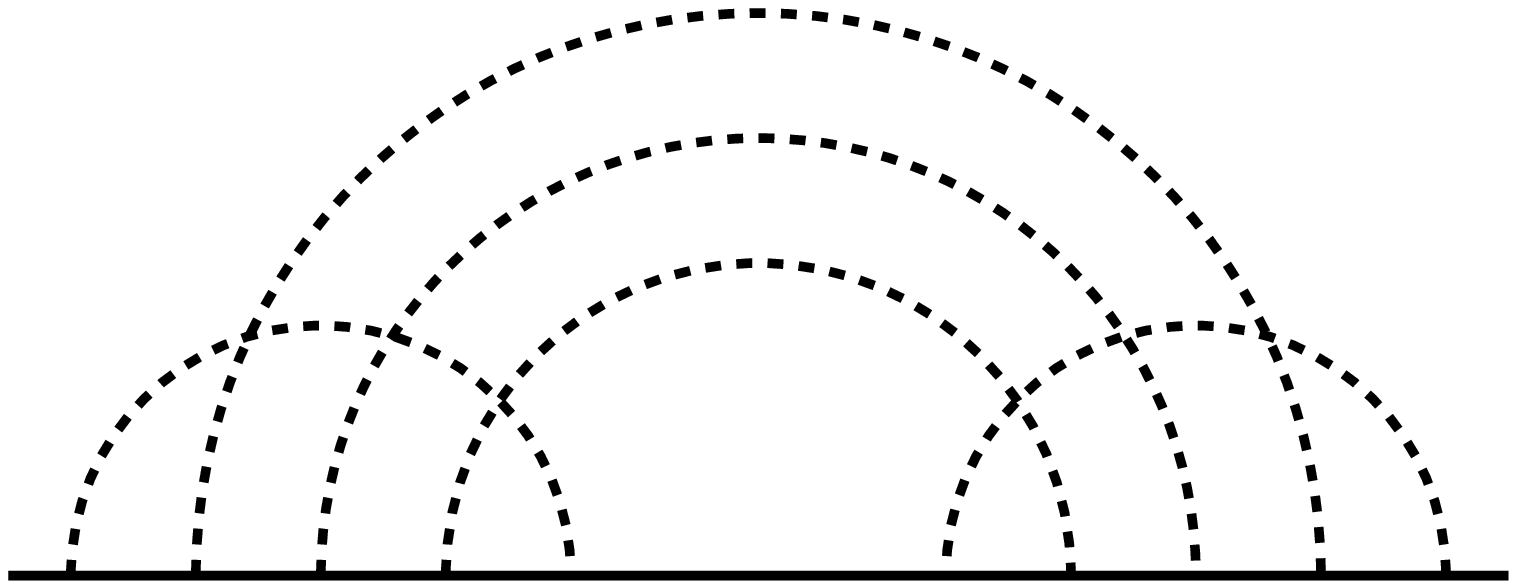} }} 
Fig.~9: A contribution of order $1/N^2$ generated by the modified recursion relation
\end{figure}

These equations are adequate to order $1/N^2$, but not to order $1/N^4$.

We summarize the steps of the new recursion relation.
\begin{itemize}
\item Assume the partition functions $Z(i,j)$ are known for all pairs
$(i,j)$ such that $i-j < l$.

\item Calculate all the one-particle irreducible functions 
$Z^{1PI}(m,n)$ in the Hartree approximation. This is just the sum of
all rainbow diagrams between $m$ and $n$, with an interaction $V_{mn}$
joining $m$ and $n$.
If no gluon is connected to the site $i$, then this
contributes to $\Gamma_{mn}^i$. 

\item Insert this function $\Gamma$ and all the functions $Z(i,j)$ in
(\ref{recursion2}) to calculate the partition functions with one more
base.
\item Iterate the process.
\end{itemize}

This procedure allows obviously to evaluate the free energy of a given RNA sequence
recursively. Regard $Z(m,n)$ as the element in the $m$th row and $n$th
column of a matrix. We impose the boundary conditions $Z(j,j)=1$ and $%
Z(j,j-1)=1$. We then use (\ref{recursion2}) to expand the matrix to ever
larger size, moving ``towards the northeast.'' In numerical evaluation, we
no longer need to know the origin of the parameter $1/N^{2}$: we can simply
take $N=1$. The factor $1/N^2$ has just allowed us to extract the most
relevant diagrams beyond the Hartree theory.

To find the ``ground state configuration'' for a given RNA sequence we
simply write (\ref{recursion2}) for $Z(1,L)$ 
\begin{eqnarray}
Z(1,L)&=&Z(1,L-1)+\sum_{j=1}^{L-1}V_{jL}\{Z(1,j-1)Z(j+1,L-1)  \nonumber \\
&+& \sum_{m,n}Z(1,m-1)\Gamma _{mn}^{j}Z(n+1,L-1)\}
\end{eqnarray}
and evaluate it ``backwards''. We replace $Z(1,L)$ by the largest term on
the right hand side 
\begin{eqnarray}
Z(1,L) &\simeq& \max_{j,m,n},\biggl \{ Z(1,L-1),\ V_{jL} Z(1,j-1)Z(j+1,L-1),
\nonumber \\
&&V_{jL} Z(1,m-1)\Gamma _{mn}^{j}Z(n+1,L-1)\}\biggr\}
\end{eqnarray}

The largest term, in turn, comprises $Z$ of lower order, for which we can
apply this bactracking algorithm. Repeating this process, we obviously
obtain the dominant configuration. 

In fact, since the lowest energy configuration obtained in this way is
not necessarily feasible in real space, a better strategy
could be
to use the backtracking algorithm to generate a set of lowest energy
configurations, and check which one can be
realized with real molecules with their rigidity and chemical constraints.
For example, configurations such as the one of fig.~10 with crossing
``gluon'' lines should be discarded, as they are forbidden by steric
constraints.

\begin{figure}[tbh]
  \epsfxsize=0.4\linewidth
  \centerline{\hbox{ \epsffile{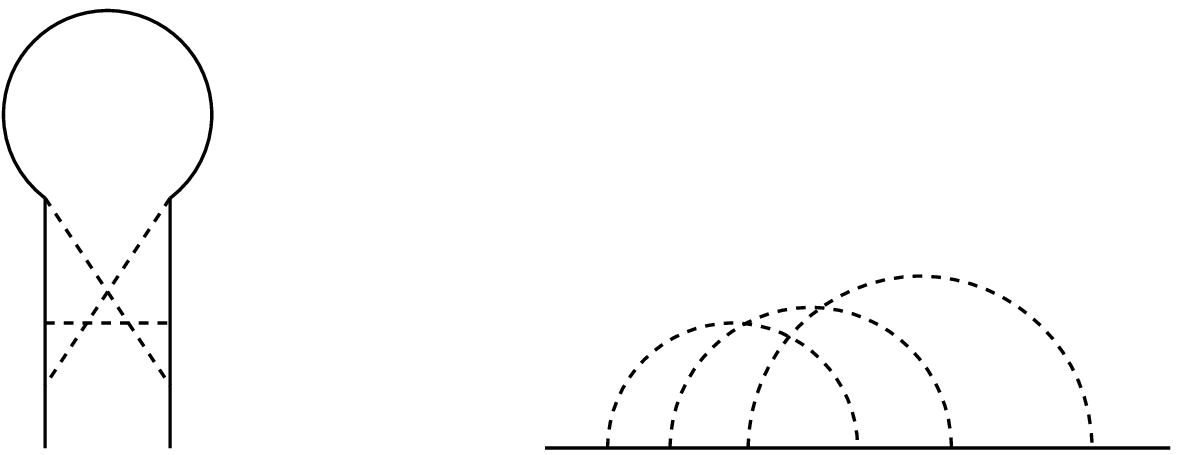} }} 
\centerline{Fig.~10: A contribution of order $1/N^2$ sterically forbidden.}
\end{figure}

\section{Conclusion}

We have shown that the RNA folding problem can be mapped onto a large $N$
matrix field theory. The dominant term ($N$ independent) is the usual
Hartree theory, which is known to generate secondary structures. The $1/N$
correction term vanishes, and the correction of order $1/N^{2}$ generates
the pseudo-knots or tertiary structure. The standard Hartree recursion relation is then
replaced by a corrected recursion relation. The resulting three dimensional
structure can be obtained by backtracking the recursion relation. The
spatial feasibility of this tertiary structure (which remains to be checked)
is due to the fact that the $1/N$ expansion classifies diagrams in terms of
their topology. What remains to be done is to include the loop entropy,
stacking energies and a numerical study of the recursion equations to order $%
1/N^{2}$, together with the backtracking algorithm. This will be presented
in a forthcoming paper.

\acknowledgments
We wish to thank Dan Holz for help with computer programming and 
Walter Fontana, Paul Higgs, Terry Hwa and Luca Peliti for very useful
comments.

This work was carried out during the ``Statistical Physics and Biological
Information'' program at the ITP, Santa Barbara, and is supported in part by
the NSF under grant number PHY 89-04035 at ITP.

\newpage

\section*{Appendix A}

In this appendix, we show that the $1/N$ correction to the free energy
vanishes identically. We first note that eq. (\ref{res1}) can be recast in
the form

\begin{eqnarray}
Z(1,L) &=&\frac{1}{N}\frac{\partial }{\partial h}\int dA_{ll^{\prime
}}^{\ast }dA_{ll^{\prime }}e^{-N\sum_{l<l^{\prime }}{\rm tr}A_{ll^{\prime
}}A_{ll^{\prime }}^{\ast }+N{\rm tr}\log M(A_{ll^{\prime }})}\left|
_{h=0}\right.  \nonumber \\
&=&\int dA^{\ast }dA\,dA_{ll^{\prime }}^{\ast }dA_{ll^{\prime }}\,A^{\ast
}e^{-N\sum_{l<l^{\prime }}\left( {\rm tr}A_{ll^{\prime }}A_{ll^{\prime
}}^{\ast }+{\rm tr}AA^{\ast }\right) +N{\rm tr}\log M(A_{ll^{\prime }},A)}
\label{newform}
\end{eqnarray}
where

\[
M(A_{ll^{\prime }},A)=\left( 
\begin{array}{lllllll}
1 & 0 & 0 & \cdot & \cdot & 0 & A \\ 
-1 & 1+a_{12} & a_{13} & \cdot & \cdot & a_{1L} & 0 \\ 
a_{12}^{\ast } & -1 & 1+a_{23} & \cdot & \cdot & a_{2L} & 0 \\ 
\cdot & \cdot & \cdot & \cdot & \cdot & \cdot & \cdot \\ 
\cdot & \cdot & \cdot & \cdot & \cdot & \cdot & \cdot \\ 
\cdot & \cdot & \cdot & \cdot & -1 & 1+a_{L-1L} & 0 \\ 
a_{1L}^{\ast } & a_{2L}^{\ast } & \cdot & a_{L-2L}^{\ast } & a_{L-1L}^{\ast }
& -1 & 1
\end{array}
\right) 
\]

The steepest descent method applied to (\ref{newform}) yields

\begin{eqnarray*}
A &=&0 \\
A^{\ast } &=&M_{L+1,1}^{-1}
\end{eqnarray*}
whereas the definition for all the other $A_{ll^{\prime }}$ and $%
A_{ll^{\prime }}^{\ast }$ are identical to those of section IV and V. The
correction of order $1/N$ to eq.(\ref{zex}) can be easily recast in the form

\begin{eqnarray}
Z^{(1)} &=&\int da_{ll^{\prime }}^{\ast }da_{ll^{\prime }}da^{\ast }da\,\exp
\left( -{\rm tr}a_{ll^{\prime }}a_{ll^{\prime }}^{\ast }-{\rm tr}aa^{\ast }-%
\frac{1}{2}{\rm tr}\left( M_{0}^{-1}c\right) ^{2}\right)  \nonumber \\
&\times &\left\{ A^{\ast }\left( \frac{1}{4}{\rm tr}\left(
M_{0}^{-1}c\right) ^{4}+\frac{1}{18}\left( {\rm tr}\left( M_{0}^{-1}c\right)
^{3}\right) ^{2}-\frac{1}{3}a^{\ast }{\rm tr}\left( M_{0}^{-1}c\right)
^{3}\right) \right\} _{L+1,1}  \label{app1}
\end{eqnarray}
with the notations of section IV and V and $M_{0}$ denotes the matrix $M$
evaluated at the stationary point. It is clear that $a^{\ast }$ occurs only
in the term ${\rm tr}aa^{\ast }$ of the first line and in the term $a^{\ast }%
{\rm tr}\left( M_{0}^{-1}c\right) ^{3}$ of the second line of (\ref{app1}).
This second term can be integrated by part in favor of $a$, to remove all
dependence on $a^{\ast }$ except in the exponent. Once it is clear that $%
a^{\ast }$ occurs only in the exponent, we recognize the holomorphic
representation of the $\delta $-function. Thus, the integration over $%
a^{\ast }$ implies that we can set $a=0$ everywhere. This being done, we see
that all the terms like $a_{12}^{\ast },\ldots ,a_{1L}^{\ast }$ and $%
a_{2L}^{\ast },\ldots ,a_{L-1,L}^{\ast }$ are present only in the exponent
(in the ${\rm tr}a_{ll^{\prime }}a_{ll^{\prime }}^{\ast }$ term). Therefore,
we can integrate them out, and the result is again a set of $\delta $%
-function which impose 
\[
a_{12}=\ldots =a_{1L}=a_{2L}=\ldots =a_{L-1,L}=0 
\]

This procedure can be carried out recursively to ``eat up'' all the $a^{\ast
}$ and $a$ , leading to the vanishing of the $1/N$ correction (\ref{app1}).

\section*{Appendix B}

\subsection{Recursion}

An alternative strategy to evaluating $Z$ recursively is by integrating out $%
\varphi _{L+1}$ in the expression for $Z(1,L+1).$ For notational simplicity,
let us define $\mu ^{2}\equiv (V^{-1})_{L+1,L+1},$ $M\equiv \varphi _{L+1}$
and $T\equiv \sum_{i=1}^{L}(V^{-1})_{L+1,i}\varphi _{i}$. Evidently, we have
to do two Gaussian integrals over $M$: 
\begin{equation}
\int dMe^{-N{\rm tr}(TM+\frac{\mu ^{2}}{2}M^{2})}=C(\mu ,N)e^{+\frac{N}{2\mu
^{2}}trT^{2}}  \label{int1}
\end{equation}
and 
\begin{equation}
\int dMe^{-N{\rm tr}(TM+\frac{\mu ^{2}}{2}M^{2})}M=-\frac{1}{\mu ^{2}}C(\mu
,N)e^{+\frac{N}{2\mu ^{2}}trT^{2}}T  \label{int2}
\end{equation}
where (\ref{int2}) is obtained by differentiating (\ref{int1}) with respect
to the matrix $T.$ Thus, after integrating out $\varphi _{L+1}$ in $Z(1,L+1)$%
, we find that the ``action'' ${\sum_{ij} }(V^{-1})_{ij}{\rm tr} (\varphi
_{i}\varphi _{j})$ has been replaced by the effective action ${\sum_{ij} }(%
\widetilde{V}^{-1})_{ij}{\rm tr}(\varphi _{i}\varphi _{j})$ where $(%
\widetilde{V}^{-1})_{ij}=(V^{-1})_{ij}-(V^{-1})_{i,L+1}\frac{1}{%
(V^{-1})_{L+1,L+1}}(V^{-1})_{L+1,j}$. It is easy to see that $\widetilde{V}$
is the $L$ by $L$ matrix obtained by crossing out the last row and column of
the $L+1$ by $L+1$ matrix $V$, as we might have expected. Putting these
steps together we obtain 
\begin{equation}
Z(1,L+1)=Z(1,L)-\frac{1}{(V^{-1})_{L+1,L+1}}\sum_{l=1}^{L}(V^{-1})_{L+1,l}<%
\frac{1}{N}{\rm tr}(\Pi _{i=1}^{L}(1+\varphi _{i}))\varphi _{l}>
\label{rec1}
\end{equation}
where $(\Pi _{i=1}^{L}(1+\varphi _{i}))$ is ordered as before. The
expectation value of a matrix $O$ constructed out of the $\varphi _{i}$'s is
defined by 
\begin{equation}
<O>\equiv \frac{1}{A(L)}\int {\Pi_k }d\varphi _{k}e^{-N\frac{1}{2}{\sum_{ij} 
}(V^{-1})_{ij}{\rm tr}(\varphi _{i}\varphi _{j})}O
\end{equation}
In other words, $Z(1,L)\equiv <\frac{1}{N}{\rm tr}{\Pi_i }(1+\varphi _{l})>$.

To evaluate $<\frac{1}{N}{\rm tr}(\Pi _{i=1}^{L}(1+\varphi _{i}))\varphi
_{l}>$ we follow the standard procedure of replacing $\varphi _{l}e^{-N\frac{%
1}{2}{\sum_{ij} }(V^{-1})_{ij}{\rm tr}(\varphi _{i}\varphi _{j})}\rightarrow
-\frac{1}{N}\sum_{k=1}^{L}V_{lm}\frac{\delta }{\delta \varphi _{k}}e^{-N%
\frac{1}{2}{\sum_{ij} }(V^{-1})_{ij}{\rm tr}(\varphi _{i}\varphi _{j})}$.
Integrating by parts, we finally obtain 
\begin{equation}
Z(1,L+1)=Z(1,L)+\sum_{k=1}^{L}V_{L+1,k}<\frac{1}{N}{\rm tr}(\Pi
_{i=1}^{k-1}(1+\varphi _{i}))\frac{1}{N}{\rm tr}(\Pi _{j=k+1}^{L}(1+\varphi
_{j}))>  \label{result}
\end{equation}

In other words, in (\ref{rec1}) we have Wick contracted $\varphi _{l}$
with $\varphi _{k}$ in the ordered product $\Pi _{i=1}^{L}(1+\varphi _{i})$.
Evidently, $\frac{1}{N}{\rm tr}(\Pi _{i=1}^{k-1}(1+\varphi _{i}))$ is to be
interpreted as $1$ for $k=1$. Similarly, $\frac{1}{N}{\rm tr}(\Pi
_{j=k+1}^{L}(1+\varphi _{j}))$ is to be interpreted as $1$ for $k=L$.

In principle,  we can extract what we need 
from this recursion relation (\ref{result}).
We emphasize that (\ref{result}) is
derived without taking the large $N$ limit and holds for finite $N$,
including $N=1$.

\subsection{Large N Expansion}

We can now perform a large $N$ expansion, giving us a systematic way of
evaluating $Z$ to any desired order of $1/N^{2}$. In the language of quantum
chromodynamics, quantities in which the indices of the matrices $\varphi
_{j} $ are summed over such as $\frac{1}{N}{\rm tr}(\Pi
_{i=1}^{k-1}(1+\varphi _{i}))$ are known as color singlet operators. It is
well known \cite{witten} that given two color singlet operators $A$ and $B$,
the expectation value factorizes to leading order in large $N:$%
\begin{equation}
<AB>=<A><B>+<AB>_{C}  \label{facto}
\end{equation}
with the connected correlation function $<AB>_{C}$ suppressed by a factor of 
$O(1/N^{2})$ relative to $<A><B>$. It is easy to see the validity of (\ref
{facto}) by drawing a few diagrams such as those in fig.~8b. Connected
correlation functions $<AB>_{C}$ have been intensively studied \cite{Bre_Zee}
in the matrix theory literature and a good deal is known about them. Thus,
we can write in (\ref{result}) 
\begin{eqnarray}  \label{bob}
<\frac{1}{N}{\rm tr}(\Pi _{i=1}^{k-1}(1+\varphi _{i}))\frac{1}{N}{\rm tr}%
(\Pi _{j=k+1}^{L}(1+\varphi _{j}))> &=&  \nonumber \\
<\frac{1}{N}{\rm tr}(\Pi _{i=1}^{k-1}(1+\varphi _{i}))><\frac{1}{N}{\rm tr}%
(\Pi _{j=k+1}^{L}(1+\varphi _{j}))>  \nonumber \\
+<\frac{1}{N}{\rm tr}(\Pi_{i=1}^{k-1}(1+\varphi _{i}))\frac{1}{N}{\rm tr}%
(\Pi _{j=k+1}^{L}(1+\varphi _{j}))>_{C}
\end{eqnarray}
We immediately recognize that first term in (\ref{bob}) as $Z(1,k-1)Z(k+1,L)$%
. By definition, the connected correlation function $Z_{C}(1,k-1;k+1,L)%
\equiv <\frac{1}{N}{\rm tr}(\Pi _{i=1}^{k-1}(1+\varphi _{i}))\frac{1}{N}{\rm %
tr}(\Pi _{j=k+1}^{L}(1+\varphi _{j}))>_{C}$ is evaluated by contracting a
matrix $\varphi _{i}$ from one of the traces to a matrix $\varphi _{j}$ from
the other trace. Thus the exact recursion relation is given by 
\begin{eqnarray}  \label{exact}
Z(1,L+1)&=&Z(1,L)+\sum_{k=1}^{L}V_{L+1,k}Z(1,k-1)Z(k+1,L)  \nonumber \\
&+& \sum_{k=1}^{L}V_{L+1,k}Z_{C}(1,k-1;k+1,L)
\end{eqnarray}

This gives an alternative representation of (\ref{recur}). Evidently, 
\begin{equation}
Z_{C}(1,k-1;k+1,L)=<M_{L+1,k+1}^{-1}M_{k,1}^{-1}>_{C}-\frac{1}{N}%
<M_{L+1,1}^{-1}M_{k,k+1}^{-1}>_{C}
\end{equation}

In principle, we can take the exact recursion relation (\ref{exact}) and
evaluate the two terms on the right hand side to any desired order in $1/N$
and thus generate, given an RNA sequence, secondary structure, tertiary
structure, ad infinitum.

\end{document}